\documentclass[%
 aip,
 amsmath,amssymb,
 reprint,%
]{revtex4-1}

\usepackage{graphicx}
\usepackage{dcolumn}
\usepackage{bm}

\usepackage[utf8]{inputenc}
\usepackage[T1]{fontenc}
\usepackage{mathptmx}
\usepackage{etoolbox}
\usepackage{braket}
\usepackage{bbold}
\usepackage{color}

\newcommand{\sign}[1]{\mathrm{sgn}(#1)}

\makeatletter
\def\@email#1#2{%
 \endgroup
 \patchcmd{\titleblock@produce}
  {\frontmatter@RRAPformat}
  {\frontmatter@RRAPformat{\produce@RRAP{*#1\href{mailto:#2}{#2}}}\frontmatter@RRAPformat}
  {}{}
}%
\makeatother

\begin{document}

\title{Controllable  non-Hermitian  Qubit-Qubit Coupling in Superconducting Quantum Circuit}

\author{Hui Wang}
\email{wanghuiphy@126.com}
\affiliation{Inspur artificial intelligence research institute,  Jinan, 250014, China}
\affiliation{Shandong Yunhai Guochuang Innovative Technology Co., Ltd., Jinan 250101, China}

 \author{Yan-Jun Zhao}
\affiliation{Key Laboratory of Opto-electronic Technology, Ministry of Education, Beijing University of
Technology, Beijing, 100124, China}

\author{Xun-Wei Xu}
\email{xwxu@hunnu.edu.cn }
\affiliation{Key Laboratory of Low-Dimensional Quantum Structures and Quantum Control of Ministry of Education,
 Key Laboratory for Matter Microstructure and Function of Hunan Province,
  Department of Physics and Synergetic Innovation Center for Quantum Effects and Applications,
   Hunan Normal University, Changsha 410081, China}

\begin{abstract}
 We propose a theoretical scheme to realize the controllable  non-Hermitian qubit-qubit coupling
by adding a high-loss resonator in tunable coupling superconducting quantum circuit.
  By  changing the  effective qubit-qubit coupling, phase and amplitude of resonator-qubit interaction, and the
 qubits' quantum states, we can continually tune the energy level attraction,   position of EP (exceptional point),
  and the nonreciprocity in the non-Hermitian superconducting circuit.
 The EPs and  non-reciprocity can affect the quantum states' evolutions and
   exchange efficiencies for two qubits  in the  non-Hermitian superconducting circuit.
   The controllable  non-Hermitian and nonreciprocal  interactions between two qubits provide a new insights and methods for exploring
    the unconventional quantum effects in superconducting quantum circuit.
\end{abstract}
\date{\today}

\maketitle

\section{Introduction}

  Non-Hermitian system possesses complex energy spectrum, and the real and  imaginary parts
 respectively label the energy levels and loss, while the starting and ending points of energy level degenerate points are known as the EPs (exceptional points) \cite{Bender1,Bender2,Bender3,Bender4,Ganainy}. The non-Hermitian coupling is essentially  dissipative type interaction which has been proposed
 to realize the entangled quantum states\cite{Zou,Qiang,Xiong}.
The system close to exceptional points contain rich unconventional physical effects,
including the ultralow driving threshold chaos\cite{You1},
 high sensitivity metrology\cite{Wei,Liu1},  high-dimensions  skin effect\cite{Kai,Liang}, chiral heat transport\cite{Cao} and so on.
The EP (exceptional point) have been studied in single ion system\cite{Ding,Zhou},
 optical microcavity\cite{Peng,Chang,Feng,Hodaei1,Hodaei2,Wei,Zyablovsky,Zhang},
and optomechanical system\cite{Hxu,Cui,Jiang,Bernier}, magnonic system\cite{hlin,guang,ikov},  hybrid quantum system\cite{You2},
cavity magnonics \cite{Harder,Yefremenko,Hxu}, single superconducting qubit\cite{Sani,Abbasi,Navarathna}, and so on.
Some theoretical work have studied the $PT$-symmetry in superconducting circuit
 with the single qubit coupling to the  resonator\cite{Starkov,Naether}. However,  it is still challenging to experimentally
observe  such  non-Hermitian in multi-qubits superconducting circuit due to the difficulties in realizing
 controlled gain and loss.

\begin{figure}
\centering\includegraphics[bb=20 205 400 520, width=7.5 cm, clip]{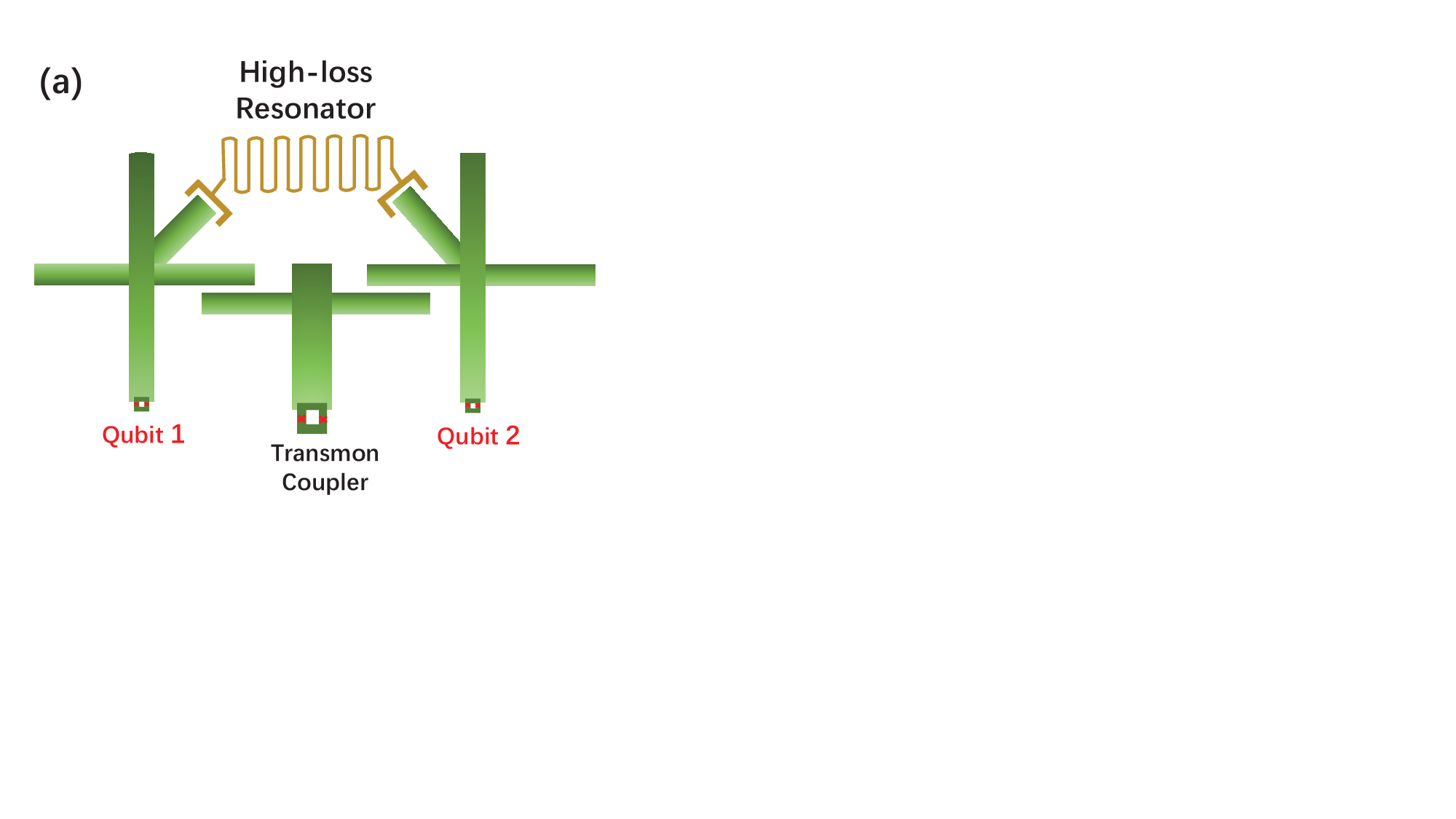}\\
\centering\includegraphics[bb=20 150 510 490, width=8 cm, clip]{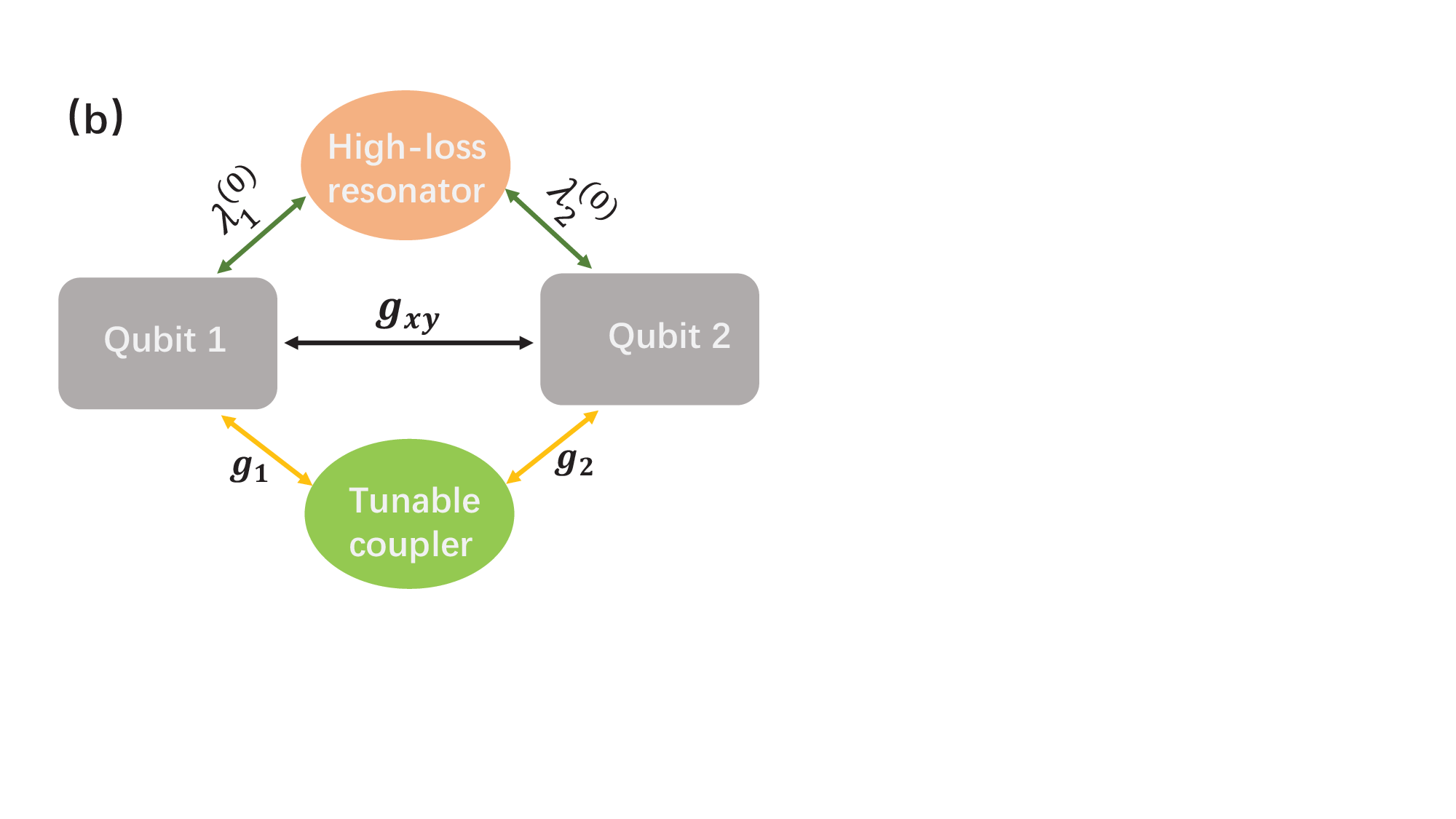}
\caption{(Color online)
  (a) Schematic diagram and (b) coupling  scheme in  superconducting quantum circuit.
 The  circuit consisting of two Xmon qubits couple to
 a common transmon coupler and a common high-loss resonator.
 $g_j$ and $\lambda^{(0)}_j$ label the respective
   coupling strengths of qubit-$j$ with the coupler and high-loss resonator,
   while  $g_{xy}$ describes the direct interaction between two qubits.
      The transmon coupler and high-loss resonator can respectively induce
      the coherent and non-Hermitian interactions between two qubits.
 }
\label{fig1}
\end{figure}

Unlike the efforts to improve the gain and loss of superconducting resonator, we use a  high-loss resonator to  supply
the effective  non-Hermitian interactions between two qubits in  tunable coupling superconducting circuit.
The effective energy levels and damping rates of two qubits can be controlled by
   effective qubit-qubit coupling,    phase and amplitude of  qubit-resonator coupling,
  and the quantum state of qubits. Because of special parameter dependent relation in superconducting circuit, the energy levels attraction and  positions of  EPs are controllable.  The non-reciprocal interactions between two qubits can also appear in  non-Hermitian superconducting quantum circuit,
  which can be used to develop the integrated unidirectional transmission devices including the microwave circulator and isolator.

 The qubits' quantum state  evolutions and states exchanges are also affected by the EPs which might be useful for the quick initialization
of qubits and the tunable quantum gates. The Rabi oscillation periods of two  non-reciprocal coupling qubits are different which can be
measured with current experimental skills.

The paper is organized as follows: In Sec.~II, we build the physical model of  non-Hermitian superconducting circuit; In Sec.~III,
the energy level attractions and  quantum state evolutions in non-Hermitian circuit are studied; In Sec.~IV,
  we analyze the  non-reciprocal coupling and quantum state exchange between two qubits.
   We finally summarize the results in Sec.~V.

\section{Physical model}

A shown in Fig.~\ref{fig1}, the system is formed by adding a high-loss resonator on the
 tunable coupling superconducting circuit consisting of two Xmon qubits and a transmon coupler.
   The transmon coupler can induce  coherent interaction between two qubits,
 and the high-loss resonator supply the  non-Hermitian interaction.
 The Xmon qubit is a modified version of transmon qubit, and the  cross-type
capacitance could  suppress the sensitivity of the qubit to charge noises \cite{Ashhab,Koch}
  and facilitate for design of superconducting quantum chip \cite{Barends}.
   The total Hamiltonian for the superconducting circuit in Fig.~\ref{fig1}
 can the obtained as $H=H_0+H_{xy}+\sum_{j}H^{(j)}_{qc}+\sum_{j}H^{(j)}_{qr}$, and the expressions of
   free Hamiltonian  $H_0$, direct qubit-qubit interaction Hamiltonian $H_{xy}$, and
coupler-qubit-$j$ interaction Hamiltonian $H^{(j)}_{qc}$ are
 \begin{eqnarray}\label{eq:1}
H_0&=&\hbar\omega_a a^{\dagger}a+\sum^2_{j=1} \frac{\hbar\omega_j}{2} \sigma^{(j)}_z+\frac{\hbar\omega_c}{2} \sigma^{(c)}_z,\nonumber\\
H_{xy}&=& \hbar g_{xy}[\sigma^{(1)}_{+}\sigma^{(2)}_{-}+\sigma^{(2)}_{+}\sigma^{(1)}_{-}],\nonumber\\
H^{(j)}_{qc}&=&\hbar g_{j} [\sigma^{(c)}_{+}\sigma^{(j)}_{-}+\sigma^{(j)}_{+}\sigma^{(c)}_{-}].
\end{eqnarray}
The  superscripts  and subscripts  $j$ label the qubit-$1$ and qubit-$2$, respectively.
$\omega_j$, $\omega_a$, and $\omega_c$ are the respectively transition
frequency of qubit-$j$, resonant frequency of resonator's fundamental mode,
 and transition frequency of coupler.  $a^{\dagger}$ and $a$ are the
  photon creation and annihilation operators of resonator's fundamental mode,
while  $\sigma^{(j)}_z$ ($\sigma^{(c)}_z$) and $\sigma^{(j)}_{\pm}$ ($\sigma^{(c)}_{\pm}$)
 label the Pauli-Z and ladder operators of  qubit-$j$ (coupler).
  $g_{xy}$ labels the direct qubit-qubit coupling strength,
   and the interaction between resonator and transmon coupler is neglected.

By applying a detuned driving field on the superconducting qubit,
the  amplitude and  phase of resonator-qubit  interaction can be tunable\cite{Pechal1,Pechal2,Magnard}.
 In this article,  the qubits and resonator are in the dispersive coupling regimes,
  that is $\lambda^{(0)}_j \ll |\Delta_{ja}|, |\Delta_{ja}+\alpha_j|$,
 where $\alpha_j$ is the anharmoncity of  qubit-$j$.
 After injecting a single tone microwave pulse with time-dependent amplitude $\Omega_j(t)$ and phase $\theta_j(t)$,
     thus resonator-qubit-$j$  interaction  becomes
     $\lambda_j\exp{(i\theta_j)}$ \cite{Pechal1,Pechal2}.
     Thus the  Hamiltonian for amplitude and phase tunable resonator-qubit-$j$ interaction is
    \begin{eqnarray}\label{eq:2}
     H^{(j)}_{qr}=\lambda_{j}[\exp{(-i\theta_{j})}a^{\dagger}\sigma^{(j)}_{-} +\exp{(i\theta_{j})} \sigma^{(j)}_{+}a],
    \end{eqnarray}
      where
     $\lambda_j=(1/\sqrt{2}) \lambda^{(0)}_j \alpha_j |\Omega_j|/[\Delta_{ja}(\Delta_{ja}+\alpha_j)]$.
   $\lambda^{(0)}_j$ is the bare coupling strength between  the first-excited state of
 superconducting artificial atom-$j$ and the fundamental mode of resonator,
  and the corresponding frequency detuning is defined as $\Delta_{ja}=\omega_{j}-\omega_a$.

In the weak dispersive coupling regimes $g_{j}/|\Delta_{jc}|\sim 1/3$ (with $\Delta_{jc}=\omega_{j}-\omega_c$),
 the Schrieffer-Wolff transformation $U=\exp{\sum^2_{j=1}(g_j/\Delta_{jc})[\sigma^{(c)}_{+}\sigma^{(j)}_{-}-\sigma^{(j)}_{+}\sigma^{(c)}_{-} ]}$
 can be applied  to cancel the qubit-coupler interaction terms,
thus the effective Hamiltonian becomes \cite{Bravyi,Richer}
\begin{eqnarray}\label{eq:3}
   H^{\prime}  & =& \hbar\omega_a a^{\dagger}a+\hbar\sum^2_{j=1}\lambda_{j}\bigg[ \exp{(-i\theta_{j})}a^{\dagger}\sigma^{(j)}_{-} +h.c.\bigg]\nonumber\\
 &+ &\hbar\bigg(\omega_{j}+\sum^2_{j=1}\frac{g^2_{j}}{\Delta_{jc}}\bigg)\frac{\sigma^{(j)}_{z}}{2}+ \hbar g_{e}\left[\sigma^{(1)}_{+}\sigma^{(2)}_{-}+h.c.\right].\quad
\end{eqnarray}
  $g_{e}=g_{xy}+g_1 g_2/\Delta_e$ is the effective  qubit-qubit coupling of coherent type, with $2/\Delta_e=1/\Delta_{1c}+1/\Delta_{2c}$.
 $g_{xy}$ describes the direct qubit-qubit coupling strength,  while $g_1 g_2/\Delta_e$ labels their indirect interaction induced by the transmon coupler\cite{Yan,Sung,Hui1,Hui2}.
     The transmon coupler is dispersively coupled with the qubits and resonator and stays in the ground state in most of the time, after neglecting
     the free Hamiltonian  the  coupler usually functions as  a tunable parameter of qubit-qubit coupling\cite{Yan}.
     These  approximations might neglect some processes of  state leakages and crosstalks related to the coupler,
     but these errors  can be significantly suppressed by the wave shape correction skill\cite{Sung}.
    During the Schrieffer-Wolff transformation, there  should be also corrected terms to qubit-resonator coupling  that is proportional to  $(\lambda_j g_j/\Delta_j) \exp{(i\theta_j)}\sigma^{(j)}_{z}(a^{\dagger}\sigma_{-}^{(c)}+\sigma_{+}^{(c)}a)$ for the first-order correction.
   The resonator and coupler are not driven and they usually stay in the ground states,
  considering their weak coupling strength and large frequency detuning,
   so the corrections to  $H^{(j)}_{qr}$ are  neglected in this article.
Since $g_{xy}\ll g_{j}$, the invariance assumption is made for $H_{xy}$ during the Schrieffer-Wolff transformation\cite{Yan}.

\begin{figure}
\centering\includegraphics[bb=2 0 500 475, width=4.26 cm, clip]{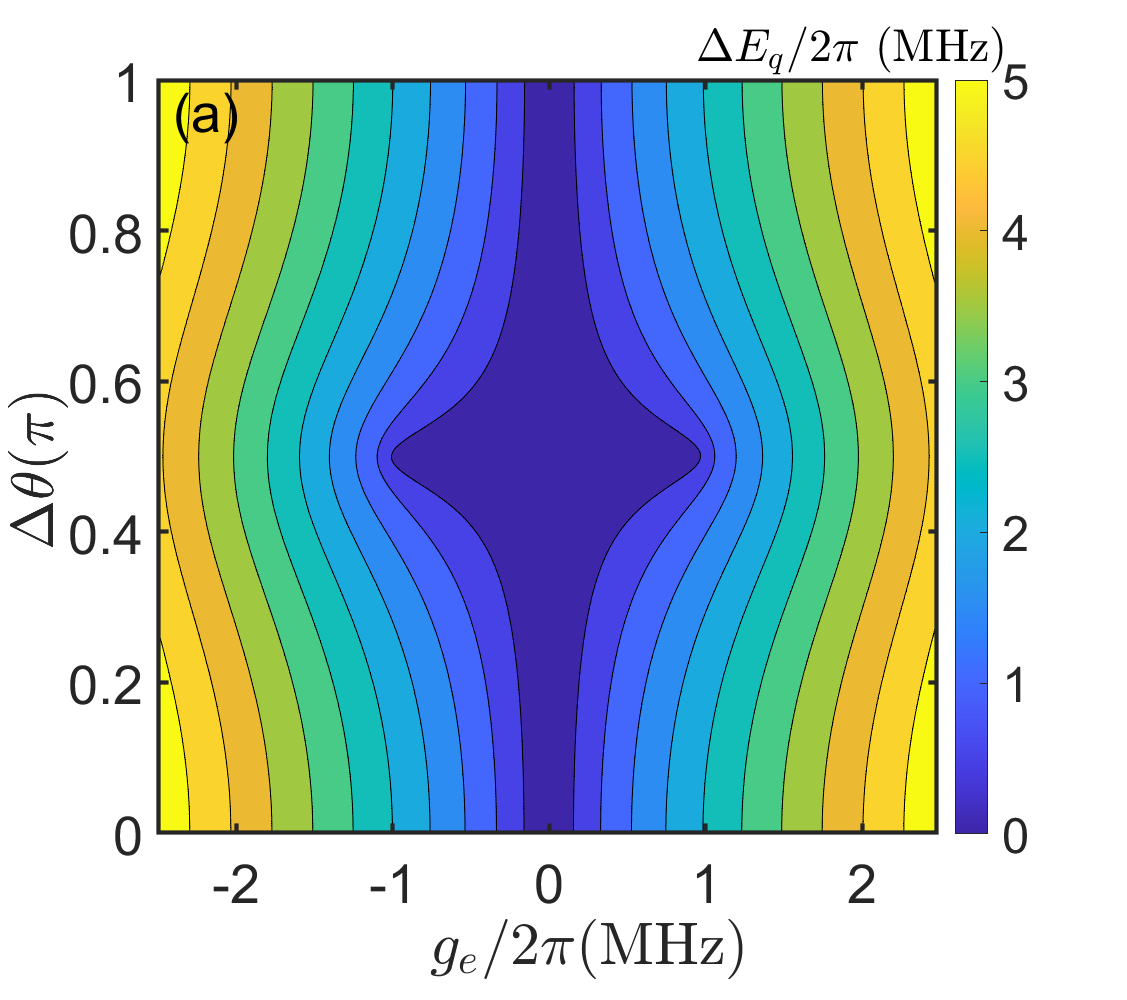}
\centering\includegraphics[bb=2 0 500 475, width=4.26 cm, clip]{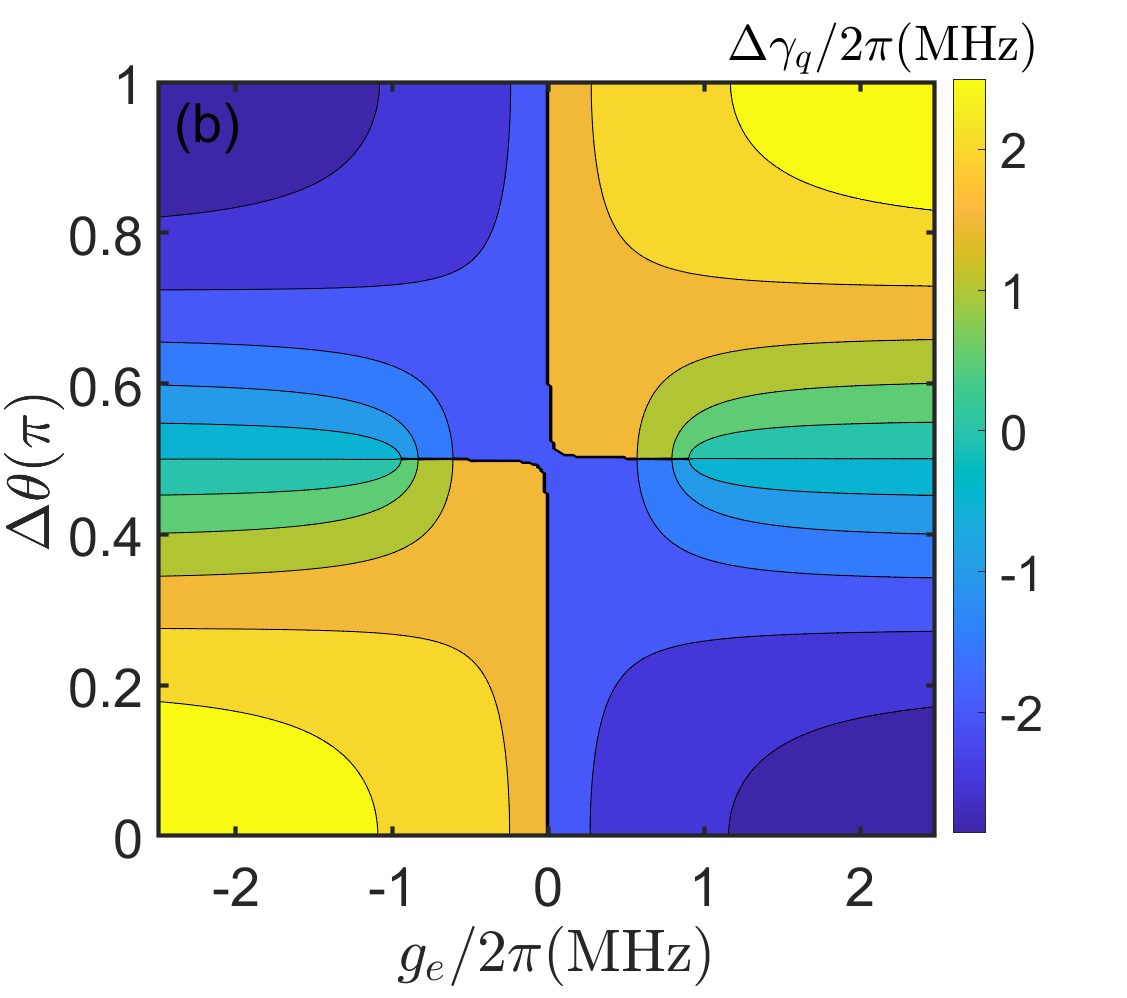}
\caption{(Color online)
 Differences in two eigeomodes' (a) energy levels ($\Delta E_q$) and (b) damping rates ($\Delta \gamma_q$)
  in the parameter space  $\{g_e,\Delta\theta\}$.
Both two qubits  in the ground states: $\langle \sigma^{(j)}_z\rangle=-1$,  with $j=1,2$.
 The other  parameters: $\omega_a/2\pi=4.475$ GHz, $\omega_1/2\pi=4.5$ GHz,
    $\omega_2/2\pi=4.505$ GHz, $\omega^{(max)}_c/2\pi=5.20$ GHz,
  $\gamma_1/2\pi=1.00$ MHz, $\gamma_2/2\pi=1.01$ MHz, $g_{xy}/2\pi=4.0$ MHz,
   $\lambda_{1,2}/2\pi=11$ MHz,
  $g_{1}/2\pi=30$ MHz, $g_{2}/2\pi=30.3$ MHz, and $\gamma_a/2\pi=65$ MHz.}
\label{fig2}
\end{figure}

In the rotating coordinates at resonator's  frequency, the  equations of motion for  qubit-$j$ and photon in resonator are
\begin{eqnarray}\label{eq:4}
 & & \dot{a}= -\gamma_{a} a-i\sum^2_{j=1}\lambda_{j}\exp{(-i\theta_{j})}\sigma^{(j)}_{-},\nonumber\\
& &  \dot{\sigma}^{(j)}_{-}= -\left[\frac{\gamma_j}{2}+i\left(\Delta_{ja}+\frac{g^2_{j}}{\Delta_{jc}}\right)\right]\sigma^{(j)}_{-}\nonumber\\
& &-i\lambda_{j}\exp{(i\theta_{j})}\sigma^{(j)}_{z}a-ig_{e}\sigma^{(j)}_{z}\sigma^{(j+1)}_{-},
\end{eqnarray}
where  $\Delta_{ja}=\omega_{j}-\omega_a$, and we assumed $j+1=1$ for $j=2$.
In the case of $\gamma_a \gg \gamma_j$,  with the adiabatic approximation, we can get
$a\approx  -i\sum^2_{j=1}\lambda_{j}\exp{(-i\theta_{j})}\sigma^{(j)}_{-}/\gamma_{a}$.
Substituting it into the motion equation of $\sigma^{(j)}_{-}$, thus we get
\begin{eqnarray}\label{eq:5}
\dot{\sigma}^{(j)}_{-}&= &  -\bigg[\big(\frac{\gamma_j}{2}+\frac{\lambda^2_{j}\sigma^{(j)}_{z}}{\gamma_{a}}\big)+i\big(\Delta_{ja}+\frac{g^2_{j}}{\Delta_{jc}}\big)\bigg]\sigma^{(j)}_{-}\nonumber\\
&-& \bigg[ig_{e}+\frac{\lambda_{j} \lambda_{j+1}\exp{[i(\theta_{j}-\theta_{j+1})]}}{\gamma_{a}}\bigg]\sigma^{(j)}_{z}\sigma^{(j+1)}_{-}.
\end{eqnarray}
 Pauli-Z operator $\sigma^{(j)}_{z}$ describes the population of qubit-$j$ and can be controlled by external driving fields.
Though coupling to a high-loss resonator, the qubit' decoherence time  should be much longer than the single qubit
  operation time, and the coherent time can be greatly enhanced by the tantalum-based superconducting devices\cite{Place,Wang11}.
The quantum states of qubits are  controlled by the driving field, and the $\langle\sigma^{(j)}_z\rangle$ can
  be considered as a transient stability value in the short period of experimental measurements.
 Thus the effective  Hamiltonian for non-Hermitian superconducting circuit is
\begin{eqnarray}\label{eq:6}
\frac{H_{non}}{\hbar}=\bigg(
           \begin{array}{cc}
             \frac{1}{2}(\Delta^{\prime}_{1a}-i \Gamma_{1}) & g_{e}-\frac{i \lambda_{1} \lambda_{2}\exp{(i\Delta\theta)}}{\gamma_{a}}  \\
             g_{e}-\frac{i \lambda_{1} \lambda_{2}\exp{(-i\Delta\theta)}}{\gamma_{a}}  & \frac{1}{2}(\Delta^{\prime}_{2a}-i\Gamma_{2}) \\
           \end{array}
         \bigg),
\end{eqnarray}
where $\Delta\theta=\theta_{1}-\theta_{2}$, $\Delta^{\prime}_{ja}=\Delta_{ja}+g^2_{j}/\Delta_{jc}$,
 and $\Gamma_{j}=\gamma_j+\lambda^2_{j}\langle\sigma^{(j)}_{z}\rangle/\gamma_{a}$.
 The  eigenmodes of $H_{non}$ are
\begin{eqnarray}\label{eq:7}
 \omega_{\pm}=\frac{\Delta^{\prime}_{1a}+\Delta^{\prime}_{2a}}{4}-i\frac{\Gamma_{1}+\Gamma_{2}}{4}\pm \sqrt{R+iI},
\end{eqnarray}
with
 \begin{eqnarray}\label{eq:8}
R &=& \frac{(\Delta^{\prime}_{1a}-\Delta^{\prime}_{2a})^2}{4}-\frac{(\Gamma_{1}-\Gamma_{2})^2}{4} +4g^2_{e}- \frac{4\lambda^2_{1}\lambda^2_{2}}{\gamma^2_{a}},\\
I&=&-\frac{8 g_{e}\lambda_{1}\lambda_{2}}{\gamma_{a}}\cos(\Delta\theta) -\frac{(\Gamma_{1}-\Gamma_{2})(\Delta^{\prime}_{1a}-\Delta^{\prime}_{2a})}{2}.
\end{eqnarray}
  $Re{(\omega_{\pm})}$  and   $Im{(\omega_{\pm})}$ respectively
 describe the effective energy levels  and  damping rates of   two eigenmodes.
   $|R|$ and $|I|$ decide separation distance between two energy levels,
while $\sign{R}$  and $\sign{I}$ decide the level repulsions or attractions
and  relative size of  damping rates.

 In the article, the  analytical and numerical calculations will adopt  the experiment accessible parameters  which have been widely
used on the  Xmon-based superconducting quantum chip\cite{Arute,Sirui,Zehang,Qian}.
The  dispersive coupling between Transmon   coupler and qubits is the mainstream solution
 to tune the qubit-qubit coupling\cite{Yan,Sung}.
  The resonator can also function as a coupler to tune the coupling between two qubits,
which has been theoretically and experimentally explored\cite{Qian,Hui1,Hui2,Yang}.
 The high-loss resonator is an unusual device on superconducting
  quantum circuit which needs special design and fabrication techniques.

\section{Controllable non-Hermitian  coupling}

\subsection{Energy level  attraction and degeneracy}

The complex energy spectrum of  eigenmodes can be tuned by
 the phase and amplitude of resonator-qubit coupling,   effective qubit-qubit coupling,
 and  qubits' populations. In the case of $R<0$ and $I=0$,
   the  $\sqrt{R+iI}$ becomes a pure imaginary number,
thus the energy levels of  two eigenmodes merge as one ($\Delta E_q =0$) and  the EP can appear.
  Even for a small  nonzero value of $I$, the  energy levels of  two eigenmodes can not overlap,
  and thus the EPs should be absent\cite{Bernier}.

\begin{figure}
\centering\includegraphics[bb=0 0 585 415, width=8.8 cm, clip]{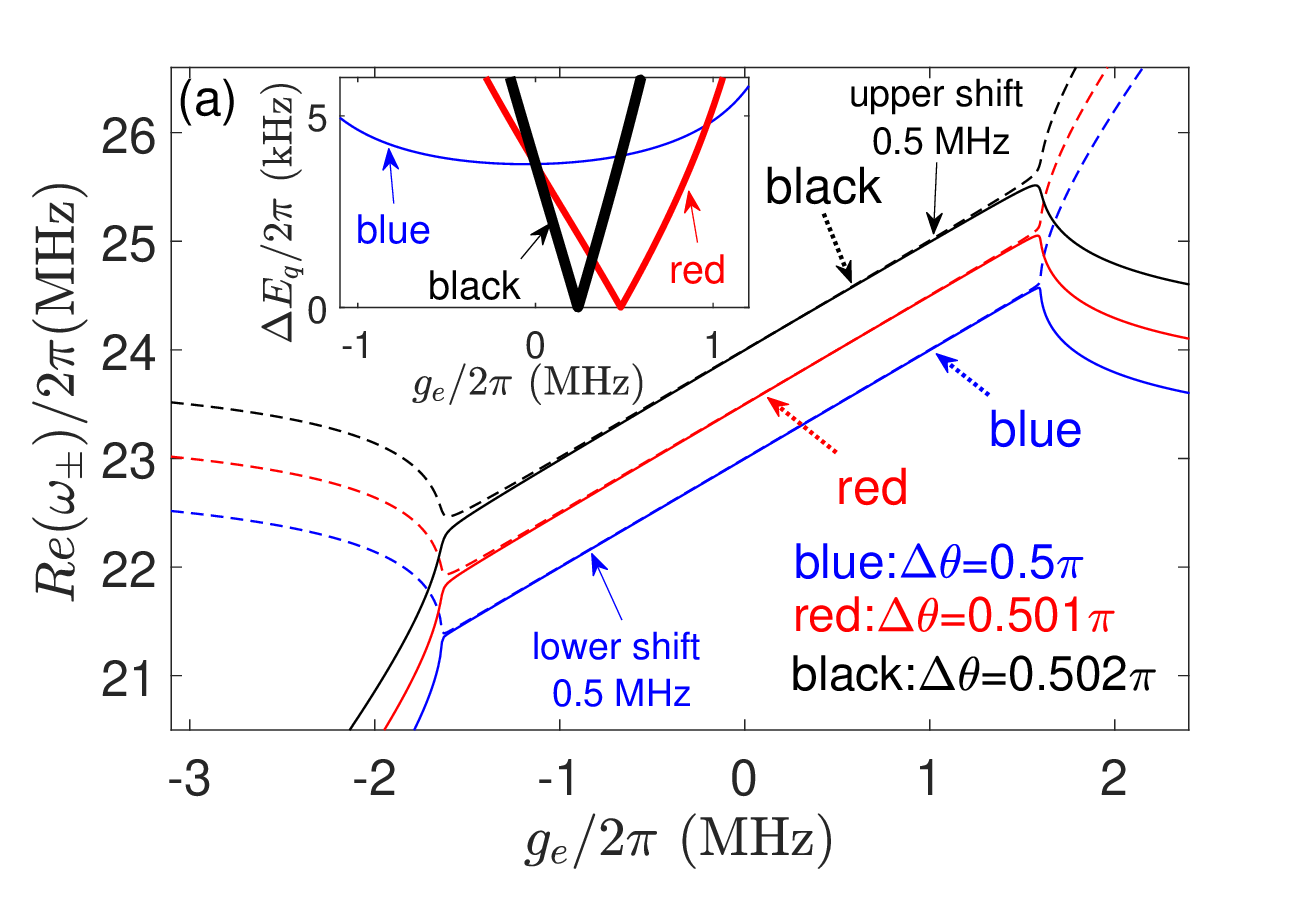}\\
\centering\includegraphics[bb=0 2 585 415, width=8.8 cm, clip]{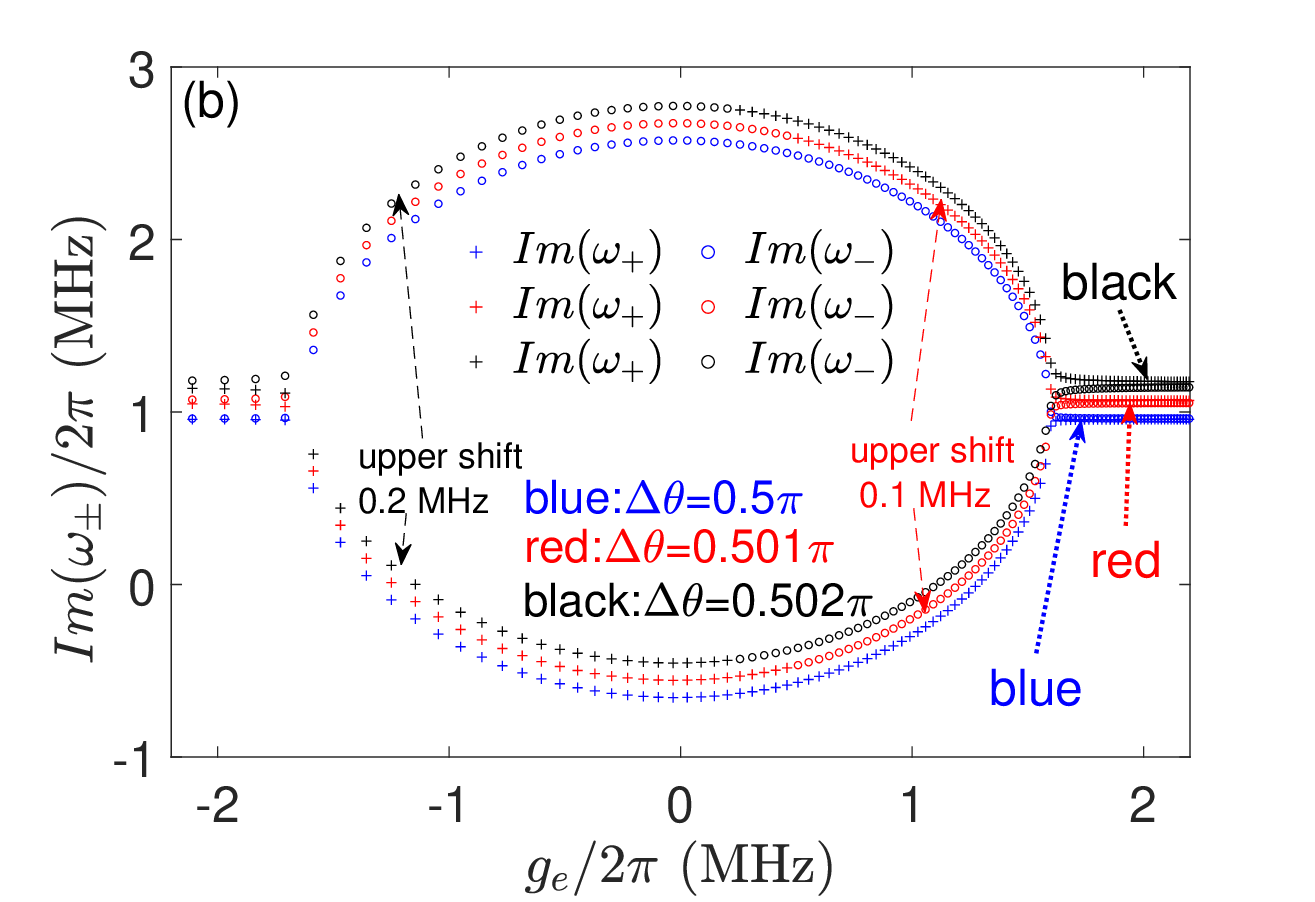}
\caption{(Color online) Energy level attraction. Eigenmodes'
 (a) energy levels $Re{(\omega_{\pm})}$  and (b)  damping rates $Im{(\omega_{\pm})}$
  as the function of $g_e$. The inset image show the energy level difference  $\Delta E_q$ of two Eigenmodes.
Three pairs of colored curves in each figure for $\Delta\theta=$:
$0.5\pi$ (blue curves); $0.501\pi$ (red curves); and $0.502\pi$ (black curves).
  Both two qubits  in the ground states: $\langle \sigma^{(j)}_z\rangle=-1$,  with $j=1,2$.
   $\lambda_{1}/2\pi=11.3$ MHz, $\lambda_{2}/2\pi=11.6$ MHz, and the
  other parameters are the same as in Fig.~\ref{fig2}.
}
\label{fig3}
\end{figure}

The differences in energy levels or damping rates of   two eigenmodes can be defined as
 $\Delta E_{q} = Re{(\omega_{+})}-Re{(\omega_{-})}= 2 Re{(\sqrt{R+iI})}$ or
$\Delta \gamma_{q} = Im{(\omega_{+})}-Im{(\omega_{-})}\approx 2Im{(\sqrt{R+iI})}$, respectively.
 In the case of $\langle\sigma^{(j)}_z\rangle =-1$ ($\Gamma_1-\Gamma_2\neq 0$),
 the  $\Delta E_q$  and  $\Delta\gamma_q$
 as the functions of  $g_e$ and $\Delta\theta$
  are plotted in Figs.~\ref{fig2}(a) and \ref{fig2}(b), respectively.
As shown  in Fig.~\ref{fig2}(a),  $\Delta E_q$ gets the local minimal values at  deep blue color  regimes.
The local  minimal values of $\Delta \gamma_{q}$ locate at the regimes close to
 $g_e/(2\pi) \approx\pm 1$ MHz  and $\Delta\theta\approx\pi/2$
as shown in Fig.~\ref{fig2}(b).
 Thus EPs should be found  close to the  regimes of $\Delta E_q \rightarrow 0$  and  $\Delta\gamma_q \rightarrow 0$ which  corresponds to
  $|g_e|/(2\pi)\approx \pm 1$ MHz  and $\Delta\theta\rightarrow\pi/2$.

\begin{figure}
\centering\includegraphics[bb=0 5 582 430, width=8.8 cm, clip]{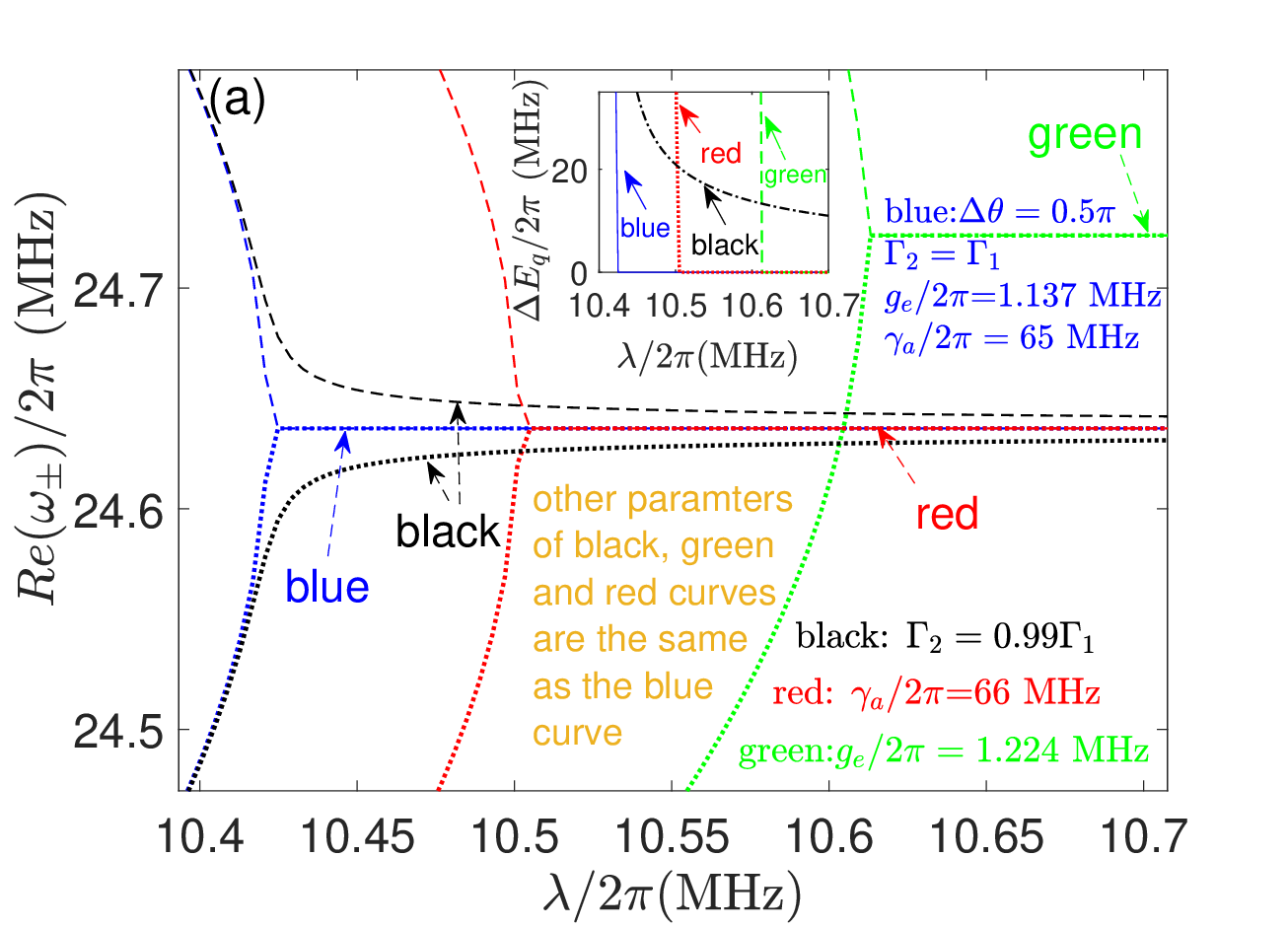}\\
\centering\includegraphics[bb=0 5 582 430, width=8.8 cm, clip]{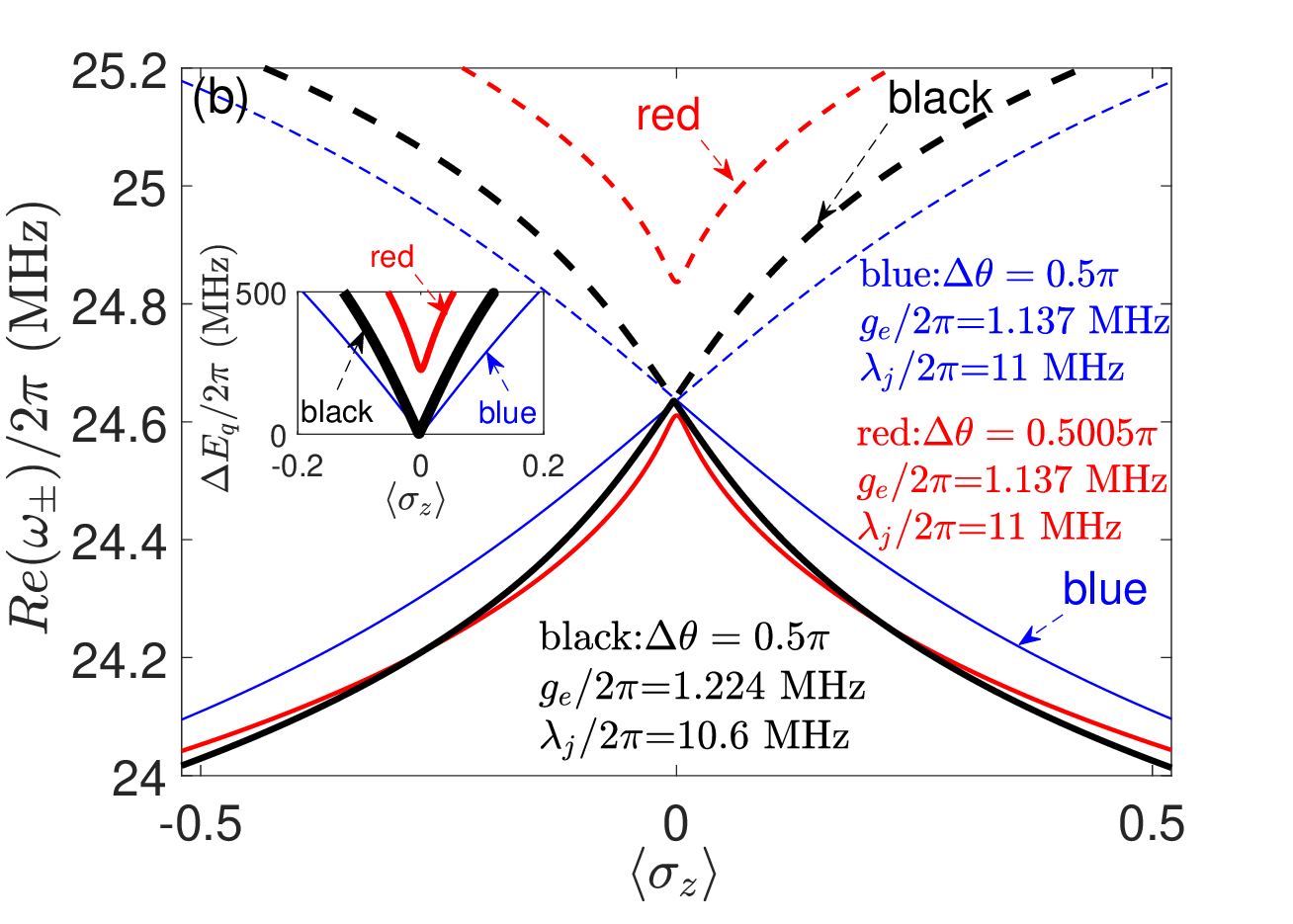}
\caption{(Color online)  Controllable Energy levels and EPs.
   Controllable Energy levels and EPs.
   $Re{(\omega_{\pm})}$ as the function of (a) $\lambda$ (with $\lambda=\lambda_{1}=\lambda_{2}$) and
   (b) $\langle \sigma_z\rangle$ (with $\langle \sigma_z\rangle=\langle \sigma^{(1)}_z\rangle=-\langle\sigma^{(2)}_z\rangle$).
 The parameters for the pair of curves in blue color in (a) for:
 $\Delta\theta=0.500\pi$, $\Gamma_2=\Gamma_1$, $g_e/2\pi=1.137$ MHz, and
   $\gamma_a/2\pi=65$ MHz. Other curves take the same parameters of blue-colored curves
   except: $g_e/2\pi=1.224$ MHz for green-colored curves;
 $\gamma_a/2\pi=66$ MHz for red-colored curves, and $\Gamma_2=0.99 \Gamma_1$ for black-colored curves.
   The three pairs of colored curves in (b) for:
  $\Delta\theta=0.5\pi$, $\lambda_j/2\pi=11$ MHz, and $g_e/2\pi=1.137$ MHz (blue curves);
  $\Delta\theta=0.5005\pi$, $\lambda_j/2\pi=11$ MHz, and $g_e/2\pi=1.137$ MHz (red curves);
  $\Delta\theta=0.5\pi$, $g_e/2\pi=1.224$ MHz, and $\lambda_j/2\pi=10.6$ MHz (black curves).
  The inset figures in (a) and (b) label the corresponding level difference ($\Delta E_q)$.
  Two qubits are assumed in the ground states in (a), and the other parameters are the  same as in Fig.~\ref{fig2}.
}
\label{fig4}
\end{figure}

 By setting  two qubits in the ground states,  the two eigenmodes' energy levels
 and damping rates  as the function of  coupling strength $g_e$
  are plotted in Figs.~\ref{fig3}(a) and ~\ref{fig3}(b), respectively.
 Energy level attraction corresponds to the regimes of nearly degenerate  energy levels
   coexisting with large separation for damping rates in Fig.~\ref{fig3}.
  In order to see clearly, the pair of curves in blue and black  colors
    are respectively moved by -0.5 MHz or 0.5 MHz  parallel to the y-axis  in Fig.~\ref{fig3}(a).
 In the case of $\Delta\theta =\pi/2$, the first term in the right side of Eq.~(9) is always zero,
  so there is no level degenerate point on pair of curves in blue color (see inset image)
  since $\Delta^{\prime}_{1a}-\Delta^{\prime}_{2a}\neq 0$
  in the parameter area of drawing. In the case of $\Delta\theta\neq\pi/2$,
    it is possible to satisfy $I=0$ as the variations of $g_e$ (or $\omega_c$),
 so the level degenerate points appear on the pairs  of curves in red or black color.
   The scattering curves for damping rates of two eigenmodes are shown in  Fig.~\ref{fig3}(b),
    to see clearly the red and black colored curved are  moved by 0.1 MHz and 0.2 MHz parallel to the y-axis, respectively.
   On the pair of curves in red or black color, the eigenmodes' damping rates  exchange with each other
  at the positions corresponding to level degenerate points in Fig.~\ref{fig3}(a), so the    energy level degenerate point should be an EP.

\begin{figure}
\centering\includegraphics[bb=21 0 445 380, width=4.20 cm, clip]{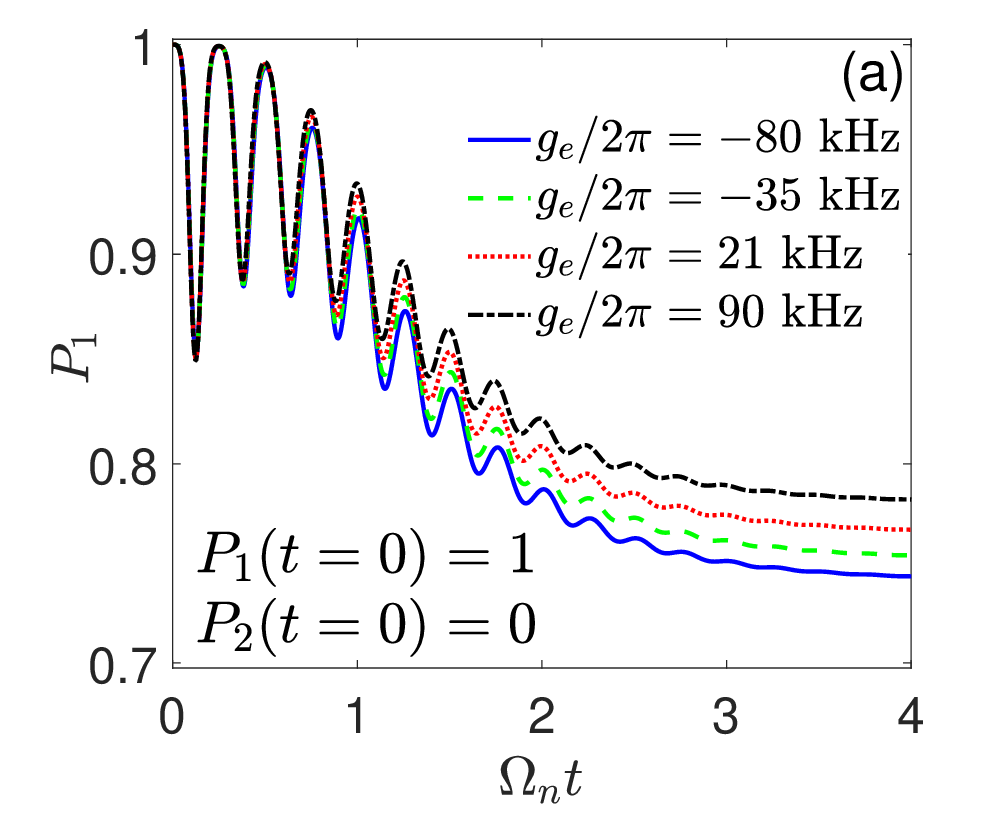}
\centering\includegraphics[bb=5 0 445 375, width=4.3 cm, clip]{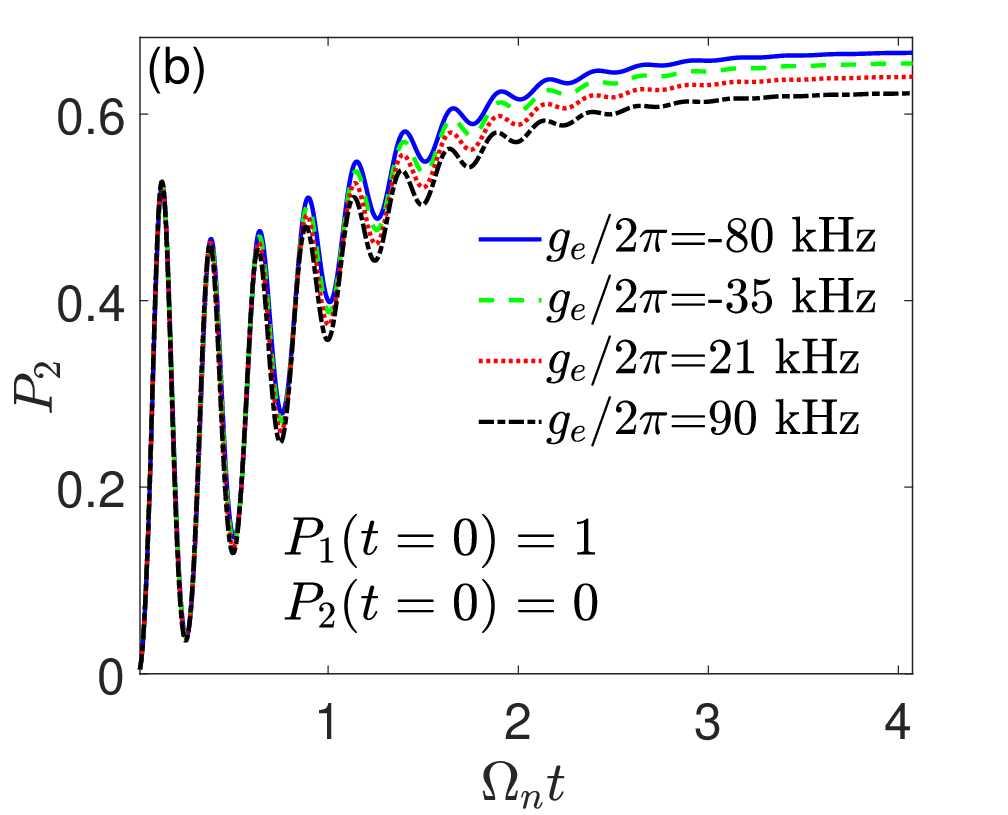}\\
\centering\includegraphics[bb=6 5 440 380, width=4.27 cm, clip]{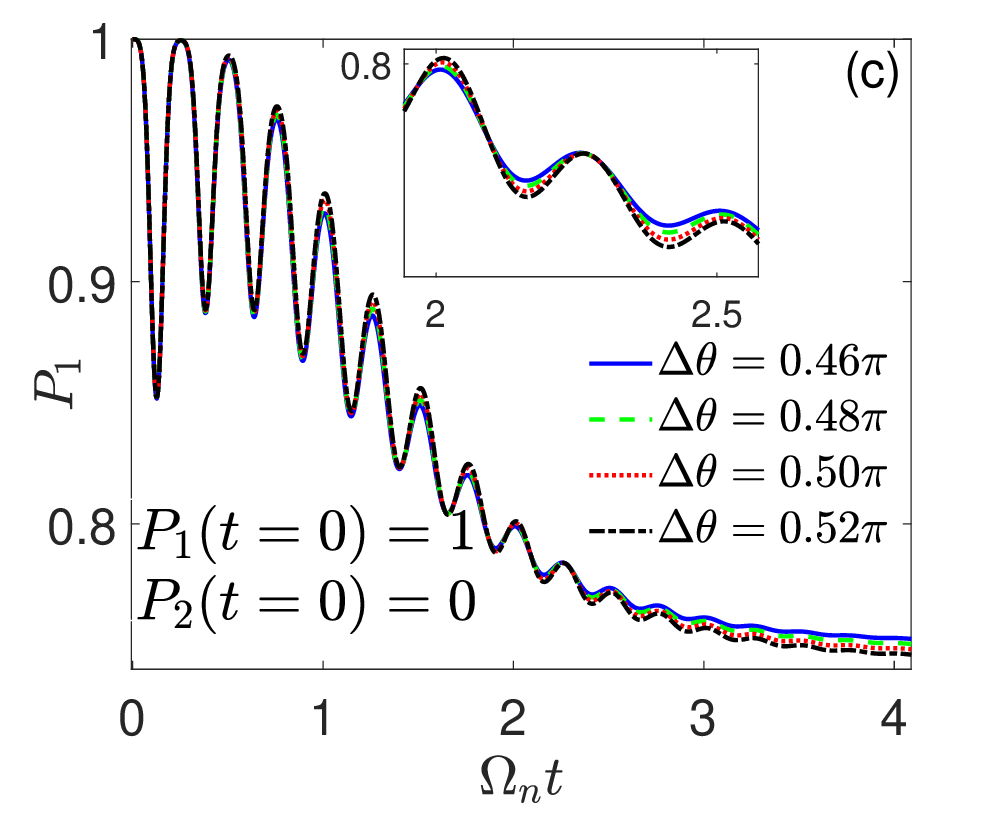}
\centering\includegraphics[bb=6 5 440 380, width=4.27 cm, clip]{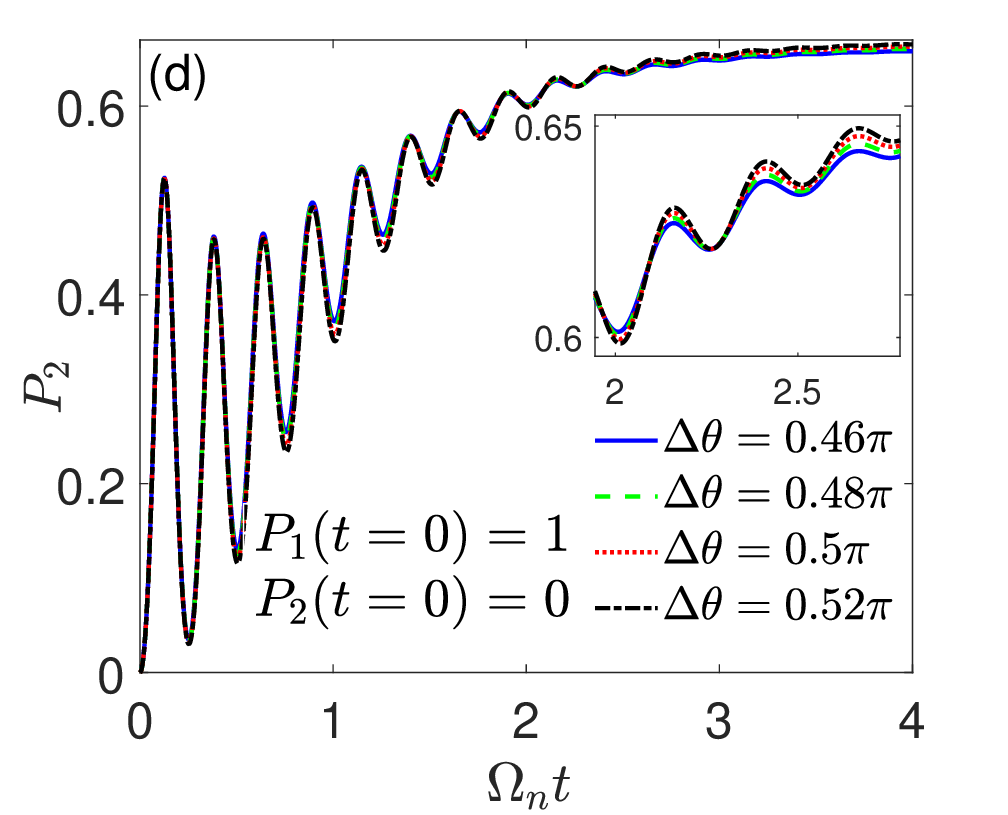}
\caption{(Color online) Controllable  quantum state evolutions.
The  occupation probabilities of  qubit-$1$ are shown in (a) and (c),
 while (b) and (d) describe the  occupation probabilities of  qubit-$2$.
The four  colored  curves in (a) and (b) for different effective
 coupling strengths $g_e/2\pi=$: -80 kHz (blue-solid curves);  -35 kHz (green-dashed curves);
21 kHz (red-dotted curves); 90 kHz (black-dash-dotted curves).  The four colored  curves in  (c) and (d) for phase difference $\Delta\theta=$ :
$0.46\pi$ (blue-solid curves);  $0.48\pi$ (green-dashed curves); $0.50\pi$ (red-dotted curves);
$0.52\pi$ (black-dash-dotted curves).
 The $\Delta\theta=5\pi/12$ in (a) and (b), and
 $g_e/2\pi=-31$ kHz in (c) and (d). $\Omega_n=\lambda_1\lambda_2/\gamma_a$,
 $\sigma^{(1)}_z(t=0)=1$,  $\sigma^{(2)}_z(t=0)=-1$,
and the other parameters are the same as in Fig.~\ref{fig2}.
  }
\label{fig5}
\end{figure}

    Figure~\ref{fig4}(a) shows the variations of two eigenmodes'  energy levels as the function of
 resonator-qubit coupling strength. For $\Delta\theta=\pi/2$ and $\Gamma_1-\Gamma_2=0$,
 the first and second terms in right side of Eq.(9) are always zeroes, so the energy level degeneracies   appear
for arbitrary values of $\lambda_j$ in the energy level attraction regimes as shown by the  blue, red, or green colors curves.
If we tune the values of $\lambda_j$ and $\gamma_a$, the positions of energy levels and   EPs
 change for the pair of curves in green (and red) color (compared with the blue color curves).  For $\cos(\Delta\theta)=0$,
in the case of $\Delta^{\prime}_{1a}-\Delta^{\prime}_{2a}\neq 0$ (parameter area of drawing)
 and $\Gamma_1-\Gamma_2\neq 0$, thus the condition of $I=0$ can not be satisfied, so
 the energy level degeneracy does not appear on the pair of curves in black color.
  The  variations of two eigenmodes'  energy levels as the function of
 qubits' populations are shown in Fig.~\ref{fig4}(b).
The energy level degeneracy appears on the pairs of curves in blue or black color (see inset image)
at certain values of  $\langle\sigma^{(j)}_{z}\rangle$,,
   but there is no energy level degeneracy on the red color curves.
  To change the transient value of   $\langle\sigma^{(j)}_{z}\rangle$,
 the qubit-$j$ should be firstly initialized to the ground states and then excited
  to the target quantum states.

\subsection{Controllable quantum state evolution}

  The changes of eigenmodes' energy levels and damping rates are
  quite significant around EPs, and the effects of  EPs
  on the evolution of qubits' quantum states should be an interesting topic worth exploring.
  By  disregarding the quantum jump terms, the  time evolutions
  of  qubits in  non-Hermitian superconducting circuit can be calculated by
 the semiclassical master equation\cite{Abbasi,Roccati1,FMinganti}
   \begin{eqnarray}\label{eq:9}
    \dot{\rho}=-i(H_{non}\rho-\rho H^{\dagger}_{non}),
    \end{eqnarray}
     where $\rho$ is the density operator of non-Hermitian circuit.
      Through the numerical calculations, the time evolution of qubits'
      quantum states can be obtained.

   As indicated by  Fig.~\ref{fig2},  the EPs should appear close to
 the  regimes of $\Delta\theta\rightarrow\pi/2$ and $g_e/(2\pi)\rightarrow \pm 1$ MHz.
 By preparing the qubit-$1$ in the excited state and qubit-$2$ in the ground state,
  the time evolution  of occupation probabilities $P_j$ (j=1,2) are plotted
    in  Figs.~\ref{fig5}(a) and ~\ref{fig5}(b) (with $\Omega_n=\lambda_1\lambda_2/\gamma_a$), respectively.
      The four curves correspond to
     different effective coupling strength $g_e$, and the nearly periodic oscillation on the curves
     indicate the energy exchanges between two qubits.
     The state exchange efficiency between two qubits
    becomes weaker close to  EPs on the black dash-dotted curves,
    where the non-Hermitian interaction  enhance the damping rates of two qubits.
    As the time evolves, the  envelopes of curves for occupation  probability $P_1(t)$ ($P_2(t)$)
      decrease (increase) to certain values.

 The phase difference $\Delta\theta$ also affects the qubits' occupation probabilities
 as shown in  Figs.~\ref{fig5}(c) and ~\ref{fig5}(d). The  envelopes of $P_1(t)$ ($P_2(t)$)
   increase (decrease)  at the beginning and   finally  stabilize at certain values.
    The variation of  relative sizes between  $\sigma^{(1)}_z$ and $\sigma^{(2)}_z$
   can change the value and even  sign of $(\Gamma_1-\Gamma_2)$ in Eqs.~(8) and (9).
This lead to the change for the sorts of colored curves, and thus the envelopes of four curves
 experience separation, polymerization, and they finally separate and  stabilize at different values.

  \begin{figure}
\centering\includegraphics[bb=0 2 440 390, width=4.28 cm, clip]{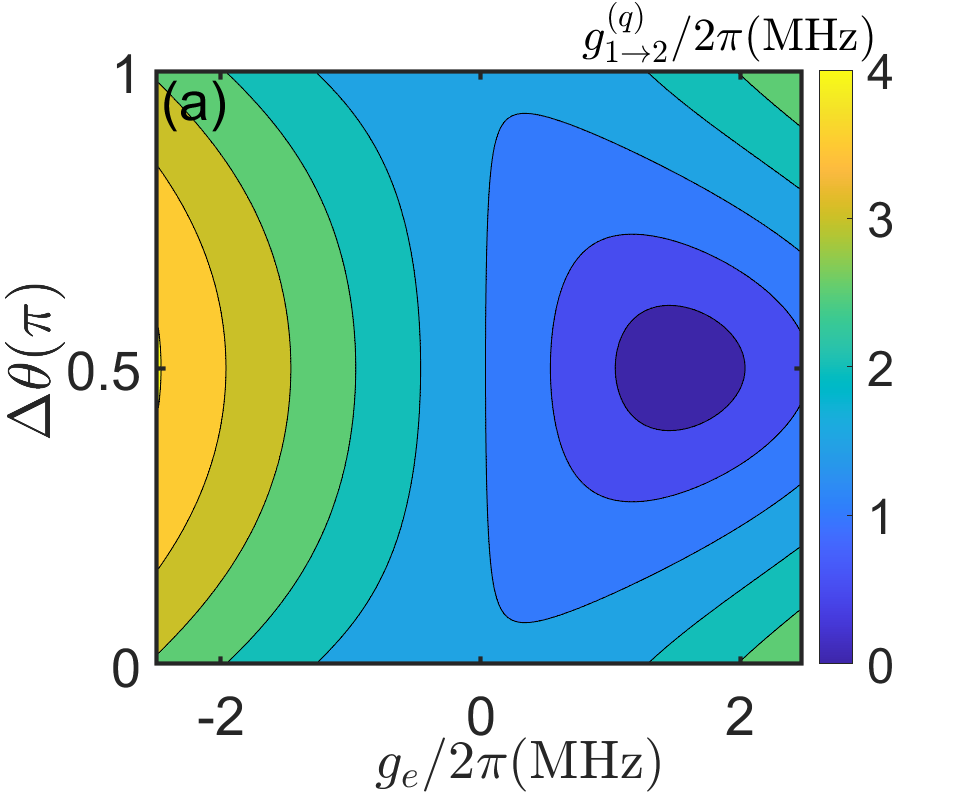}
\centering\includegraphics[bb=0 5 440 385, width=4.28 cm, clip]{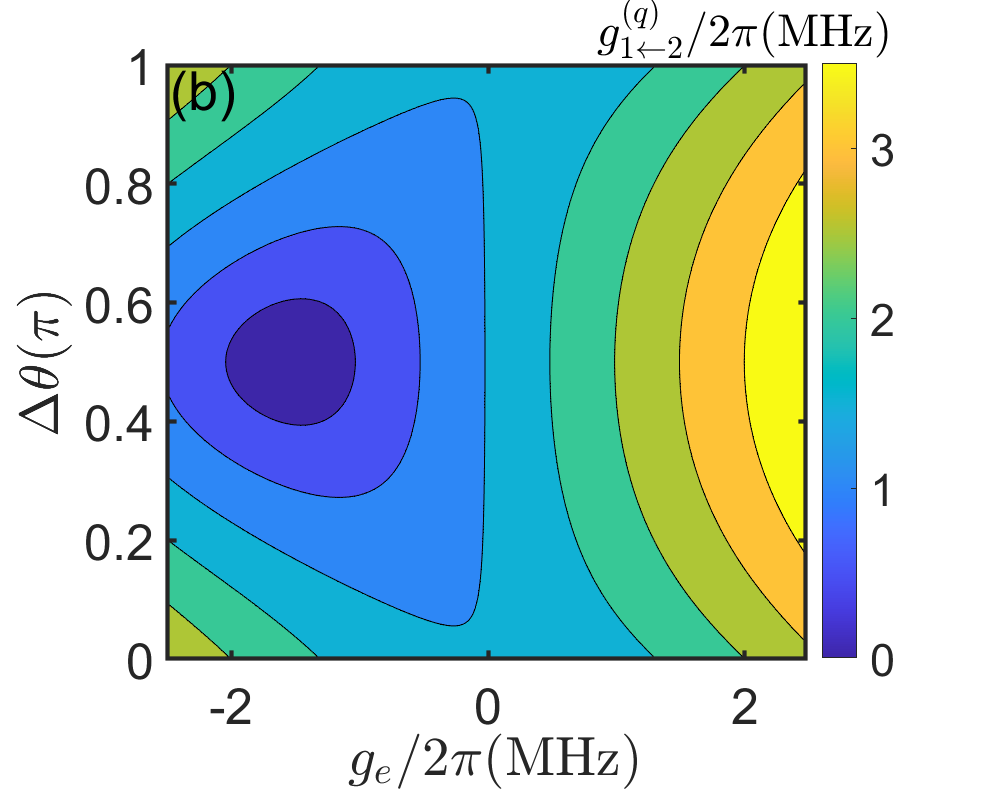}\\
\centering\includegraphics[bb=6 5 490 450, width=4.28 cm, clip]{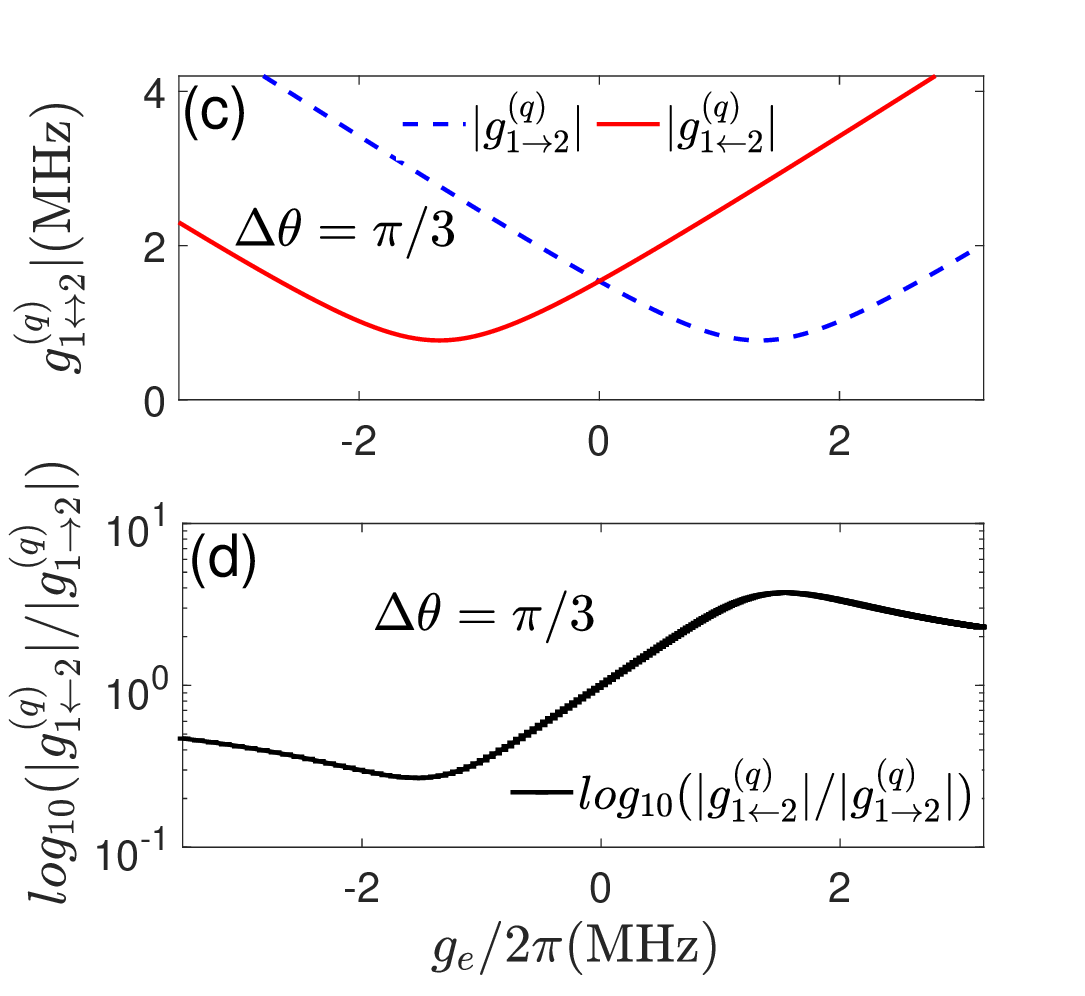}
\centering\includegraphics[bb=6 8 490 450, width=4.28 cm, clip]{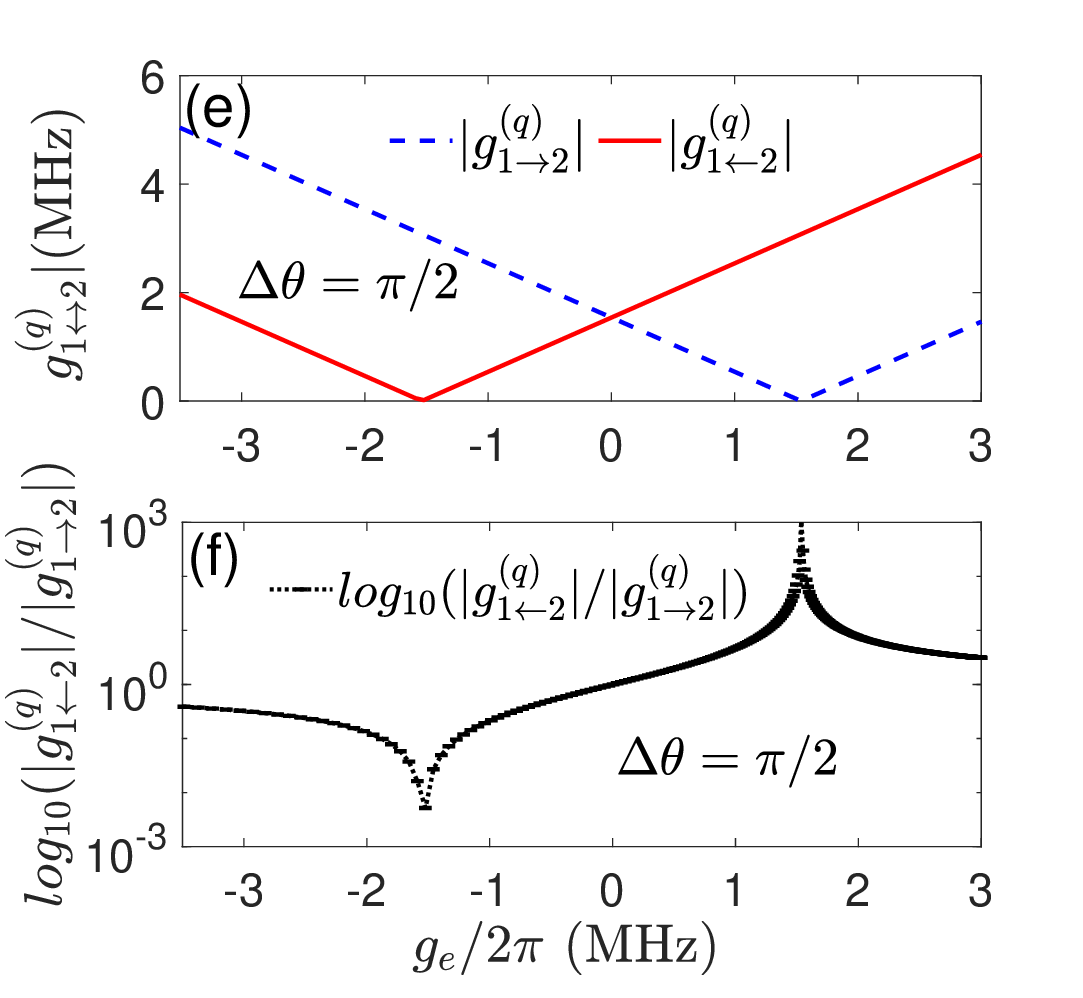}
\caption{(Color online) Nonreciprocal qubit-qubit coupling.
  Pseudo-color images of  (a) $g^{(q)}_{1\rightarrow2}$ and (b) $g^{(q)}_{1\leftarrow2}$ in the  parameter space $\{\Delta\theta, g_e\}$.
  One dimensional curves of  $g^{(q)}_{1\leftrightarrow 2}$  as the function of $g_e$ for
  (c) $\Delta\theta=\pi/2$ and (e)  $\Delta\theta=\pi/3$,
 and  corresponding non-reciprocal ratios $|g^{(q)}_{1\leftarrow2}|/|g^{(q)}_{1\rightarrow 2}|$
  at (d) $\Delta\theta=\pi/2$  and (f) $\Delta\theta=\pi/3$.
 The other parameters are:  $\omega_a/2\pi=4.475$ GHz, $\omega_1/2\pi=4.5$ GHz,
    $\omega_2/2\pi=4.505$ GHz, $\omega^{(max)}_c/2\pi=5.20$ GHz,
  $\gamma_1/2\pi=1.00$ MHz, $\gamma_2/2\pi=1.01$ MHz, $g_{xy}/2\pi=4.0$ MHz,
   $\lambda_{1,2}/2\pi=11$ MHz,
  $g_{1}/2\pi=30$ MHz, $g_{2}/2\pi=30.3$ MHz, $\gamma_a/2\pi=65$ MHz,
   and  $\langle \sigma^{(j)}_z\rangle=-1$  (with $j=1,2$).
 }.
\label{fig6}
\end{figure}

\section{Controllable Nonreciprocal  coupling}

Many studies have shown that the non-reciprocality can appear
 in the non-Hermitian systems\cite{Xu2,Xu3,Dizicheh,Huang,Roccati2,Roccati3}.
 In this section, we will explore the non-reciprocal coupling between
  two qubits in the superconducting circuit of Fig.~\ref{fig1}.
As indicated by Eq.~(6), the  bi-directional  qubit-qubit coupling  in non-Hermitian circuit
 can be  defined as $g^{(q)}_{1\rightleftharpoons2}=g_{e}-i \Omega_n\exp{(\mp i\Delta\theta)}$, respectively.
We can still get  $g^{(q)}_{1\rightarrow 2}\neq g^{(q)}_{1\leftarrow 2}$  in the case of $\Delta\theta=n\pi$.
  The local maximal (or minimal) values  of $g^{(q)}_{1\rightarrow 2}$ and $g^{(q)}_{1\leftarrow 2}$
 don't  overlap with each other in Figs.~\ref{fig6}(a) and \ref{fig6}(b),
 which indicates the non-reciprocal coupling between two qubits\cite{Huang}.
One dimensional curves of $g^{(q)}_{1\rightleftharpoons2}$ as the function of coupling strength $g_e$
are plotted in  Figs.~\ref{fig6}(c) ($\Delta\theta=\pi/3$) and  ~\ref{fig6}(e) ($\Delta\theta=\pi/2$),
 and the corresponding curves of nonreciprocal ratio $|g_{1\leftarrow2}|/|g_{1\rightarrow 2}|$
  are shown  in  Figs.~\ref{fig6}(d) and ~\ref{fig6}(f), respectively.
The  different locations  for local minimal (or maximal) values of $g^{(q)}_{1\rightarrow 2}$
 and $g^{(q)}_{1\leftarrow 2}$  indicate the asymmetric
 qubit-qubit  interactions,  and the degree of non-reciprocity can be tuned by the phase difference $\Delta\theta$.

\begin{figure}
\centering\includegraphics[bb=0 7 535 480, width=4.28 cm, clip]{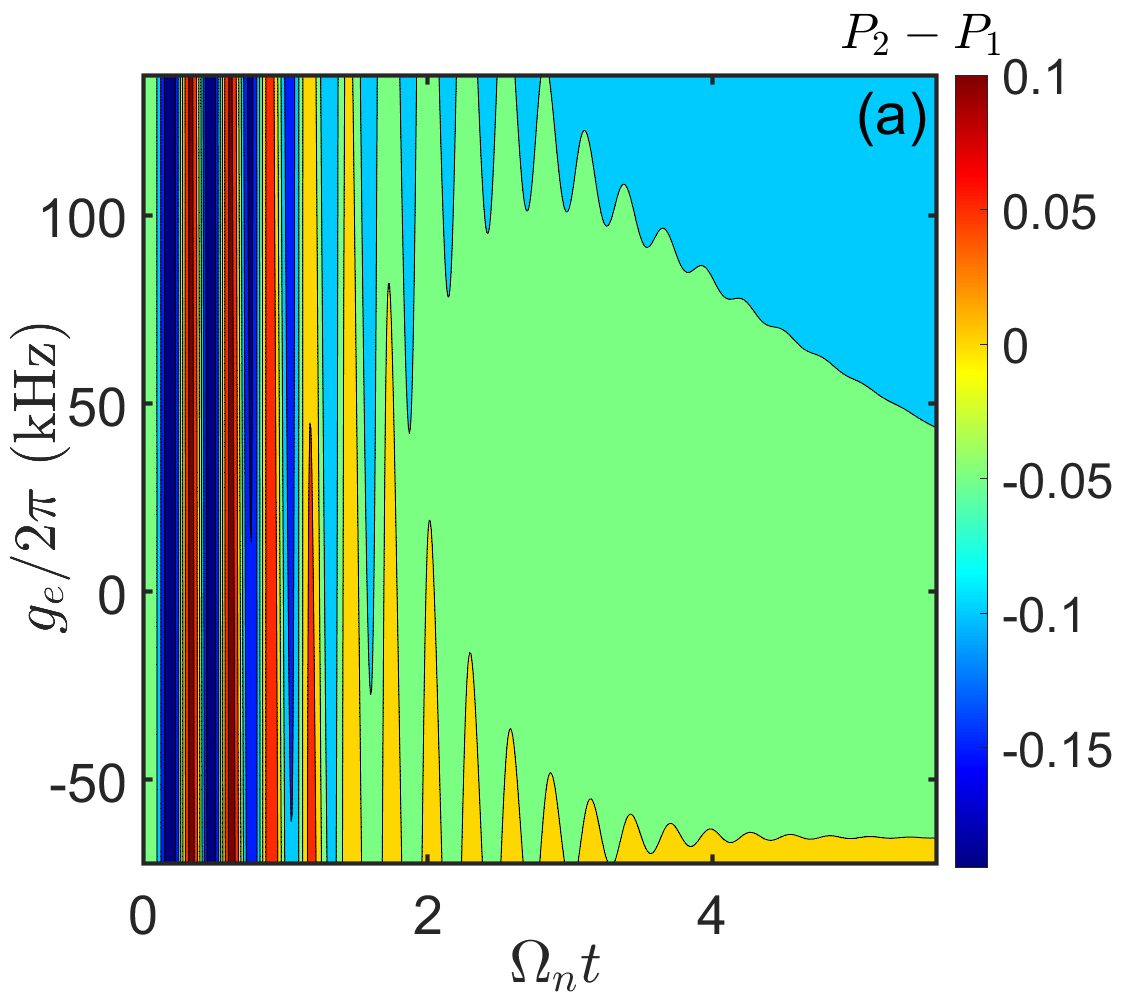}
\centering\includegraphics[bb=0 8 535 480, width=4.28 cm, clip]{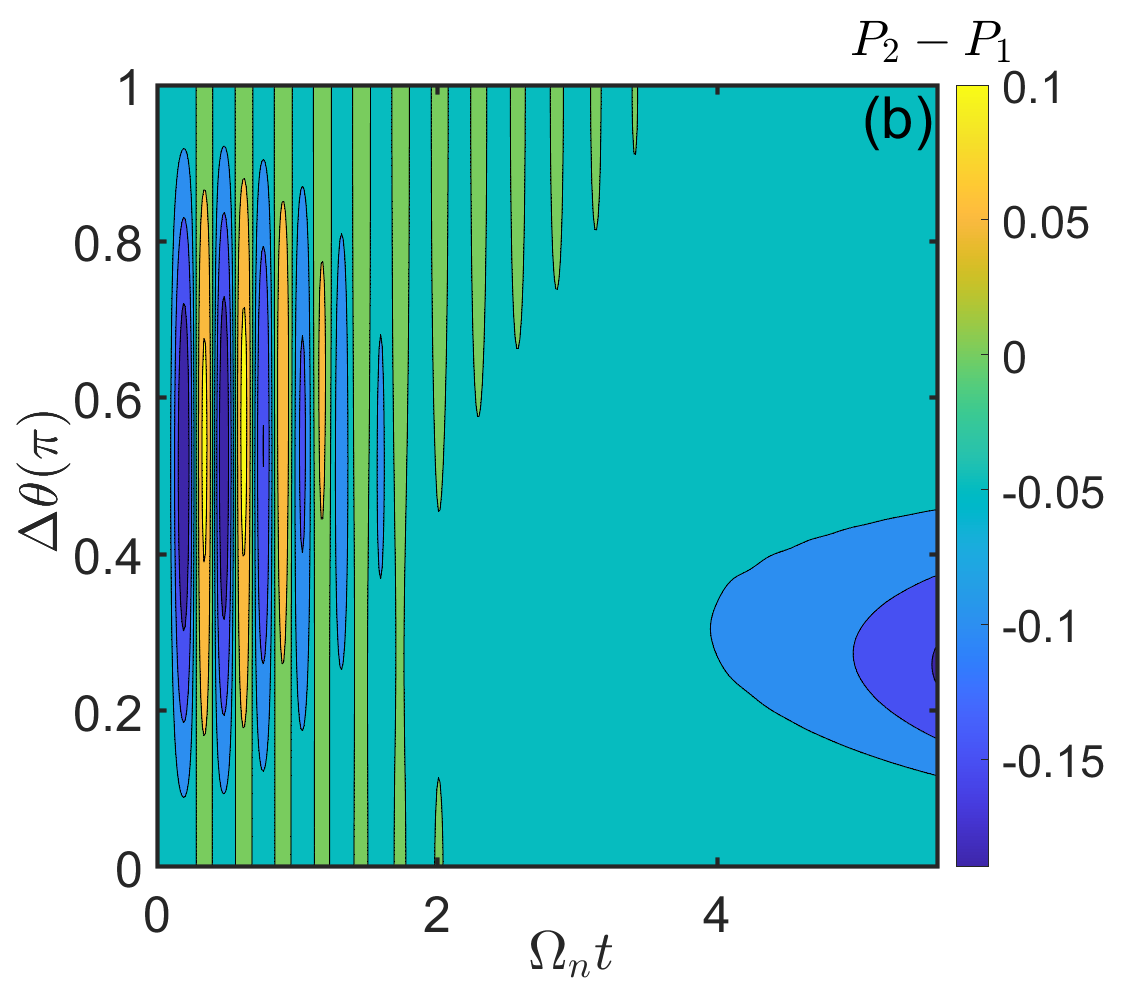}
\caption{(Color online) Asymmetric quantum state evolutions.
 Pseudo-color images for the difference in occupation probabilities ($P_2-P_1$)  in the parameter spaces
  (a) $\{\Omega_n t, g_e \}$ and (b) $\{\Omega_n t, \Delta\theta \}$.
   $\Delta\theta=\pi/2$ in (a) and $g_e/2\pi=5.3$ kHz in (b), and the other parameters are:
$\gamma_a/2\pi=65$ MHz, $g_{xy}/2\pi=4.0$ MHz, $\omega_j/2\pi=4.5$ GHz,  $g_j/2\pi=30$ MHz,
   $\lambda_{j}/2\pi=11$ MHz,  $\sigma^{(j)}_z(t=0)=1$, and $\gamma_j/2\pi=1$ MHz, with $j=1,2$.
 }.
\label{fig7}
\end{figure}

  The  non-reciprocal coupling  might differentiate the time evolution processes of two identical qubits.
 By choosing the same transition frequencies, coupling strengths,  damping rates, and initial states,
    the  time evolution  processes of two qubits can be calculated with the master equation in Eq.(10).
  For the initially excited states of two qubits, the difference in the occupation probabilities
  of two  qubits are plotted in the parameter spaces of $\{\Omega_n t, g_e \}$ in
    Fig.~\ref{fig7}(a) and   $\{\Omega_n t, \Delta\theta \}$ in Fig.~\ref{fig7}(b), respectively.
    The non-zero of $(P_2-P_1)$ in some regimes of Pseudo-color images  in Fig.~\ref{fig7} reveal the asymmetry between
    the quantum state evolutions of  non-reciprocal coupling qubits.
    The near periodic stripes of  $(P_2-P_1)$ on Pseudo-color images indicate the
     different Rabi oscillation  frequencies  between two qubits, which can be easily measured with
     current low  temperature measurement skills.

\section{Conclusion}

The energy spectrum and quantum state evolutions are
studied in non-Hermitian superconducting quantum circuit.
The non-Hermitian and  non-reciprocal interactions can be used to control
the quantum state evolutions and state exchanges of two qubit,
 which might be useful for enhancing the fidelity of quantum gates and
 speeding up quantum state initialization on superconducting quantum chip.
In this manuscript, we used mainstream parameters of current
 Xmon-based tunable coupling superconducting quantum chip which can be
  easily realized by current experiment skills. One main challenge is
  the high-loss superconducting resonator, but the fabrication of resonator with
    with the normal metal might be a promising choice.

\section{Acknowledgment}

We thank Zhiguang Yan for the useful discussions. H.W. is supported the Natural Science Foundation of Shandong Province under Grant No. ZR2023LZH002 and the Inspur artificial intelligence research institute. Y.-J.Z. is supported by National Natural Science Foundation under grant No. 62474012, Beijing Natural Science Foundation under grant No. 4222064, Scholarship from China Scholarship Council, Beijing Outstanding Young Scientist Program (JWZQ20240102009).  X.-W.X. is supported by the National Natural Science Foundation of China (Grant Nos. 12064010 and 12247105), the Science and Technology Innovation Program of Hunan Province (Grant No. 2022RC1203), the Natural Science Foundation of Hunan Province of China (Grant No. 2021JJ20036), and the Hunan provincial major sci-tech program (Grant No. 2023ZJ1010).

\end{document}